\documentstyle[epsfig,12pt]{article}

\newcommand{\nc}{\newcommand}
\nc{\beq}{\begin{equation}}   \nc{\eeq}{\end{equation}}
\nc{\beqa}{\begin{eqnarray}}  \nc{\eeqa}{\end{eqnarray}}
\nc{\lsim}{\begin{array}{c}\,\sim\vspace{-21pt}\\< \end{array}}
\nc{\gsim}{\begin{array}{c}\sim\vspace{-21pt}\\> \end{array}}
\nc{\create}{\hat{a}^\dagger}   \nc{\destroy}{\hat{a}}
\nc{\kvec}{\vec{k}}             \nc{\kvecp}{\vec{k}^\prime}
\nc{\kvecpp}{\vec{k}^{\prime\prime} }   \nc{\kb}{\bf k}
\nc{\kbp}{{\bf k}^\prime}       \nc{\kbpp}{{\bf k}^{\prime\prime}
}
\nc{\bfk}{{\bf k}}              \nc{\cohak}{a_{{\bf k}}}
\nc{\cohap}{a_{{\bf p}}}        \nc{\cohaq}{a_{{\bf q}}}
\nc{\cohbk}{b^*_{{\bf k}}}      \nc{\cohbp}{b^*_{{\bf p}}}
\nc{\cohbq}{b^*_{{\bf q}}}
\nc{\PP}{{\bf P}} \nc{\HH}{{\bf H}}

\begin{document}

\title{{\large {\bf On the Intrinsic Parity of Black Holes}}}
\author{
Stephen D.H.~Hsu\thanks{hsu@duende.uoregon.edu} \\
Department of Physics \\
University of Oregon, Eugene OR 97403-5203 \\}
%\date{July, 2002}
\maketitle

\begin{abstract}
We investigate the intrinsic parity of black holes. It appears
that discrete symmetries require the black hole Hilbert space to
be larger than suggested by the usual quantum numbers M (mass), Q
(charge) and J (angular momentum). Recent results on black hole
production in trans-Planckian scattering lead to gravitational
effects which do not decouple from low-energy physics. Dispersion
relations incorporating these effects imply that the
semi-classical black hole spectrum is similar in parity even and
odd channels. This result can be generalized to other discrete and
continuous symmetries.

\end{abstract}

\newpage

\section{Black hole quantum numbers}

No-hair theorems \cite{nohair} suggest that black holes are
classified by quantum numbers such as mass M, charge Q and angular
momentum J. (There are more exotic possibilities, depending on the
details of the matter content of the model \cite{exotic}.) In this
note we investigate the intrinsic {\it parity} of black holes.

We adopt the viewpoint \cite{TH} that black holes can be described
by quantum mechanics and have a unitary S-matrix. Given this
assumption there are two possibilities: either the Hamiltonian
commutes with parity, or not. Consider the following gedanken
process. Let N massive pseudoscalar particles undergo adiabatic,
S-wave collapse to form a black hole. The parity of the initial
state is \beq \PP \vert i \rangle = (-1)^N \vert i \rangle~~~.
\eeq If semi-classical gravitational interactions are parity
invariant ($[ \PP, H ] = 0$), we have \beqa
\vert f \rangle &=& e^{-iHT} \vert i \rangle \nonumber \\\
\PP \vert f \rangle &=& e^{-iHT} \PP \vert i \rangle ~=~ (-1)^N
\vert f \rangle \eeqa The final state $\vert f \rangle$ is a black
hole characterized by its mass M, charge Q and angular momentum J
(zero in this case), but with an intrinsic parity determined by
the number of pseudoscalars N.

This suggests that the usual quantum numbers must be supplemented
by a $\pm$ parity of the black hole wavefunction under spatial
inversion $\vec{x} \rightarrow - \vec{x}$. This additional quantum
number doubles the size of the black hole Hilbert space: \beq
\vert M, J, Q \rangle ~\rightarrow~ \vert M, J, Q, \pm \rangle
~~~.\eeq

Interestingly, the Kerr metric has + parity for all angular
momenta J, whereas we expect that a generic quantum object has
parity $\pm (-1)^J$, where $\pm$ reflects intrinsic parity. Unlike
other forms of hair, intrinsic parity does not seem to have a
classical counterpart in exterior black hole solutions.

However, it isn't clear that $[H, \PP] = 0$. Consider the
evaporation of a black hole. One can make a convincing argument
that the semi-classical evaporation of a black hole violates
continuous global symmetries such as baryon number. Because there
is no classical hair associated with global symmetries, the
evaporation process is independent of global charge, and hence the
final state might have different charge than the black hole. In
the case of a continuous symmetry an arbitrary amount of charge
can be hidden in a large black hole, and the last moment of
quantum evaporation cannot compensate for a large difference
between the charges of the hole and of the semi-classical
radiation.

In the case of discrete symmetries like C, P and T this argument
is not conclusive: one could imagine each black hole storing a
single bit of information representing the net parity of
everything radiated, and then compensating for this in the final
quantum part of the evaporation. Indeed, discrete symmetries might
be the remnant of a local symmetry and therefore conserved by
gravity \cite{KW} (this is true of parity in some string models).

Let us consider the implications of $[H, \PP] \neq 0$ due to
gravitational effects. An immediate consequence is that the
Hamiltonian cannot be diagonalized on the subspace of definite
$\pm$ parity, but rather has the form \beq \label{H} H ~=~
\left(\begin{array}{ccc}
A & C^*  \\
C & B
\end{array} \right)~~~,
\eeq where A and B are real and C is complex. Each are functions
of M, J and Q. Energy eigenstates are of the form: ($|\alpha|^2 +
|\beta|^2 = 1$)
 \beqa \vert 1
\rangle &=& \alpha \vert + \rangle ~+~ \beta \vert -
\rangle \nonumber \\
\vert 2 \rangle &=& \beta \vert + \rangle ~-~ \alpha \vert -
\rangle ~~~.\eeqa Note that for every M, J and Q there are now two
energy eigenstates, generally with $E_1 \neq E_2$. The states
$\vert 1 \rangle$ and $\vert 2 \rangle$ are further
distinguishable if they decay with different probabilities to plus
or minus parity final states.

Neither $E_1$ nor $E_2$ need be equal to the classical black hole
energy as a function of M, J and Q. The part of the Hamiltonian
that violates parity must produce a mass splitting. Is there a way
to avoid these strange results? One interesting (albeit
speculative) possibility is to take \beq \label{H1} H ~=~
\left(\begin{array}{ccc}
0 & E  \\
E & 0
\end{array} \right)~~~.
\eeq so that the eigenvalues are $\pm E$. If we reinterpret the
negative eigenvalue as a kind of black hole antiparticle energy,
we obtain equal energies for each M, J and Q at the cost of
introducing a new type of (anti) black hole. The energy
eigenstates are
 \beqa \vert 1
\rangle &=& \frac{1}{\sqrt{2}} \left( \vert + \rangle ~+~ \vert -
\rangle \right) \nonumber \\
\vert 2 \rangle &=& \frac{1}{\sqrt{2}} \left( \vert + \rangle ~-~
\vert - \rangle \right)  ~~~.\eeqa and appear equally likely to
decay into plus or minus parity final states.

Whether or not $[H, \PP] = 0$, we seem to be led to a larger
Hilbert space of black hole states. Similar conclusions hold for
discrete symmetries such as {\bf C} and {\bf T}.

What are the observable consequences of intrinsic parity? Can we
give a more general argument that black holes of both parities
exist? Below we consider black hole production from the collision
of particles with known parity. Recent work reveals a range of
(trans-Planckian) energies and impact parameters where the quantum
amplitude can be reliably computed, and the cross section is
geometrical. The black holes are large compared to the Planck
length $L_{\rm Planck}$ and are themselves semi-classical.
Seemingly, they can be produced in both plus and minus parity
channels.

We investigate this using dispersion relations, which, due to
analyticity, allow us to relate low energy correlators to black
hole production at high energy. We study dispersion relations
involving correlators of opposite parity. The correlators
themselves are calculable at low energy, and through the
dispersion relations imply the existence of large, semi-classical
black holes in both $\pm$ parity channels.

Our analysis is similar in spirit to that of Peskin and Takeuchi
\cite{S}, who used precision electroweak data to constrain the
spectrum of states in technicolor models. In this case, we use low
energy physics (low means relative to the Planck scale!) to
constrain the black hole spectrum.

\section{Black hole production from collisions}

Eardley and Giddings \cite{eg} have analyzed classical solutions
in general relativity which describe the collision of
ultra-relativistic particles at non-zero impact
parameter\footnote{For previous work on high energy scattering in
general relativity, see \cite{ACV}-\cite{D'Eath:book}; for recent
work on black hole production in particle collisions, see
\cite{giddingsthomas}-\cite{SS}.}. They demonstrate the existence
of a closed trapped surface for any collision with sufficiently
small impact parameter (at fixed center of mass energy). The lower
bound on the critical impact parameter is of order the radius \beq
R_s \sim G E~~, \eeq where G is Newton's constant and E the center
of mass energy. This leads to a geometrical classical cross
section for black hole production in agreement with Thorne's hoop
conjecture \cite{hoop}.

In a classical collision which produces a black hole the maximum
angular momentum is $J_{max} \sim R_s E \sim G s$. The geometrical
cross section for fixed J is $\sigma_J (E) \sim J/ s $, and the
total cross section is \beq \sigma_{\rm total} ~=~
\int_0^{J_{max}} dJ ~\sigma_J (E) ~\sim~ J_{max}^2 / s ~\sim~
R_s^2~~. \eeq

The relationship between classical and quantum black hole
production has been investigated recently in \cite{bhp} using a
path integral representation of the S-matrix
\cite{ghp}-\cite{nato}. Although there are only two colliding
particles, there are many gravitons in the initial state due to
the strong gravitational fields. Thus, from the gravitational
point of view the initial state is semi-classical, and from this
perspective the subsequent evolution should not depend on whether
the initial two particle state is in a parity even or odd state.

One would expect that the semi-classical approximation holds as
long as $GE^2 >> 1$ and the impact parameter is larger than the
Planck length $L_{\rm Planck}$. To express this in terms of
angular momentum, note that a semi-classical initial state must
have $\Delta p << p$ and $\Delta x << x$. Hence, using the
uncertaintly principle, we have $J \sim p x >> 1$.

We can obtain a more restrictive condition by requiring that the
curvature is small everywhere up to the formation of the closed
trapped surface \cite{bhp}. This is only possible if we leave the
Aichelburg-Sexl limit of the metric, which describes an infinitely
boosted particle of zero size \cite{Aichelburg:1970dh}, and
instead consider particles of finite size r \cite{bhp,KV}. In this
case the maximum curvature in Planck units is given by \beq {\cal
R} \sim \left( {R_s \over r} \right)^2 \left( {L_{\rm Planck}
\over r} \right)^2 ~~~.\eeq The impact parameter of the collision
is given by $b \sim J/E$, and the particle size r must be less
than $b$. This leads to the requirement that $E << M_{\rm Planck}
J^{2/3}$, or equivalently that the range of validity of the
semi-classical approximation for fixed J is \beq \label{srange}
M_{\rm Planck}^2 ~<<~ s ~<<~ M_{\rm Planck}^2 J^{4/3} ~~~. \eeq
Clearly, only processes with large J can be described
semi-classically.

Let us briefly comment on regions of s outside of (\ref{srange})
for scattering at fixed J. When $s \sim M^2_{\rm Planck}$ the
resulting black hole is intrinsically quantum mechanical and
stringy effects are expected. When $s \gsim M_{\rm Planck}^2
J^{4/3}$ the resulting black hole is large and classical, however
the initial state is not semi-classical. Roughly speaking, fixing
J as $s \rightarrow \infty$ requires that the impact parameter
goes to zero. Eventually the impact parameter is smaller than the
Planck length, and the uncertainty in position $\Delta x$ is
larger than $x$. In this regime we expect black hole production,
but cannot reliably calculate its cross section. If the
geometrical result continues to hold, the dispersive integrals we
consider below will get a logarithmically divergent contribution
from production at asymptotically large s, thereby violating
unitarity. However, as discussed above, there is no reason to
expect the semi-classical result to continue to hold in this
region. We expect that the actual cross section is parametrically
smaller than the geometric one at asymptotic energies, leading to
convergent integrals.

\section{Dispersive analysis}

Analyticity allows us to relate the low- and high-energy behavior
of quantum amplitudes. For example, given a vector current
$J^{\mu}_V (x)$, we define its correlator \beq (q^\mu q^\nu - q^2
g^{\mu \nu}) \Pi_V (q^2) = i \int d^4x ~e^{iq \cdot x} \langle 0
\vert T (J^{\mu}_V (x) J^{\mu}_V (0) \vert 0 \rangle~~. \eeq Using
Cauchy's theorem and some general properties of the singularity
structure of a correlator in the complex $q^2$ plane,  we obtain
the following relation between the correlator and an integral over
its imaginary part: \beq \label{int} \Pi (q^2) = {1 \over \pi}
\int_{s_0}^\infty ds {{\rm Im} \Pi (s) \over s - q^2 - i\epsilon}
\eeq The imaginary part is also known as the spectral function,
and describes the production of resonances or multi-particle final
states with the same quantum numbers as the current.

Before proceeding further, let us discuss whether dispersion
relations like (\ref{int}) apply in the hyper-Planckian region, $s
> M_{\rm Planck}^2$. Naively, one might think that quantum gravity
governs the behavior of the integrand in this region. However, a
quantum treatment of gravity is only necessary for {\it
curvatures} larger than the Planck scale. It is easy to see that
trans-Planckian energies alone do not require quantum gravity,
since any collision of macroscopic objects (for example, billiard
balls) could involve energies larger than the Planck scale. As
discussed in the previous section, there is a region in s (given
in (\ref{srange})) in which the amplitude is dominated by
semi-classical configurations of low curvature.

More fundamentally, the derivation of (\ref{int}) only requires
unitarity and analyticity of scattering amplitudes, both
properties which we expect to hold even at high energies (for
example, in string theory). Indeed, in our use of (\ref{int}) the
left hand side is a forward scattering amplitude involving
particles of low energy and in nearly flat space time. The
relation of this amplitude to its imaginary part only requires
analytic properties which are expected to be valid everywhere in
the complex plane, even very large s. Black holes appear as poles
near the real axis in the complex s plane, and there is a branch
cut (associated with a threshold for black hole production at $s_0
\sim M_{\rm Planck}^2$) on the real axis. The part of the integral
given in (\ref{srange}) only involves semi-classical black holes
of low curvature.

Consider operators which produce states with large angular
momentum J, whose correlators receive contributions from
semi-classical black hole formation. Define $S_J$ to be the
difference of correlators of currents with opposite parity: \beq
\label{sum} S_J (q^2) = \Pi_\Gamma (q^2) ~-~ \Pi^P_\Gamma (q^2)
~~~, \eeq where we extract the dimensionless scalar function
$\Pi_\Gamma (q^2)$ from the angular momentum J component of
$\langle 0 \vert~ T ( J_\Gamma (x) J_\Gamma (0) )~ \vert 0
\rangle$. $\Pi_\Gamma (q^2)$ is a function of $q^2$ only, and we
can always rescale by $q^2$ to obtain something dimensionless. The
current is defined as $J_\Gamma = \bar{\psi} \Gamma \psi$, with
$\Gamma \sim \prod_i^J~ \hat{\partial}_i$, where $\hat{\partial}$
denotes a unit vector. We obtain the angular momentum J component
by symmetrizing and subtracting the appropriate traces. In
$\Pi^P_\Gamma$ we replace $\Gamma$ by $\gamma_5 \Gamma$.

By taking $q^2$ large and spacelike, we can ensure that short
distance contributions dominate. In an asymptotically free theory,
(\ref{sum}) can then be reliably computed, and shown to cancel up
to corrections suppressed by powers of $1 / q^2$. If there are no
low-energy condensates that appear in the operator product
expansion of the correlator, this cancellation is exact. Note, we
keep $q^2 << M^2_{\rm Planck}$, so that quantum gravity effects
remain negligible. In perturbation theory, the leading
contribution to each correlator comes from a one loop diagram. The
leading contribution to $\Pi_\Gamma (q^2)$ is of the form \beq
(-)^J \, {\rm Tr} \left[ ~S\!\!\!/ ~\Gamma^{\dagger} ~S\!\!\!/
~\Gamma~ \right]~~~, \eeq where $S\!\!\!/$ is a fermion
propagator. The leading contribution to the opposite parity
correlator $\Pi_\Gamma^P (q^2)$ is of the form \beq (-)^{J+1} \,
{\rm Tr} \left[ ~S\!\!\!/ ~\gamma_5 \Gamma^{\dagger} ~S\!\!\!/ ~
\Gamma \gamma_5 ~ \right]~~~. \eeq So, $\Pi_\Gamma (q^2) =
\Pi_\Gamma^P (q^2)$ and $S_J (q^2)$ vanishes at leading order.

However, for sufficiently large J, there will be a
trans-Planckian\footnote{Strictly speaking, correlators of
operators separated by less than a Planck length may be
ill-defined in quantum gravity. However, one can avoid this
problem by considering the dispersion relation (\ref{int}) as a
relation between scattering amplitudes and their imaginary parts,
without reference to operators or correlators. As discussed, the
scattering amplitudes of interest are well-defined even at
trans-Planckian energy.} region of s (see(\ref{srange})) in the
dispersion integral where the semi-classical approximation
applies, and the black hole production cross section is given by
$\sigma_J (s) \sim J/s$. This cross section implies constant ${\rm
Im} \Pi (s)$ in this region, with $s >> q^2$. If black holes only
exist with a single (e.g., either $+$ or $-$) intrinsic parity,
then the integrand is zero in one of the channels and the
dispersion integral yields a correction \beq \Delta S_J ~\sim~ J
\log J \eeq for large J. Since the dispersion integral (\ref{int})
runs to infinity, this result is independent of the Planck scale,
and remains even if we take $M_{\rm Planck} \rightarrow \infty$.
$\Delta S_J$ can be made arbitrarily large by considering $J
\rightarrow \infty$, which contradicts the result that (\ref{sum})
must vanish at large $q^2$.

The region $s > M_{\rm Planck}^2 J^{4/3}$ gives another unbalanced
contribution to $\Delta S_J$ if there are not black holes of both
parities, although we cannot calculate its size. It is highly
unlikely that these contributions are cancelled by stringy
processes in the region $s \sim M_{\rm Planck}^2$. This would
require that string scattering be highly parity asymmetric.
Further, the string cross sections would have to be larger than
geometric in order to cancel the contributions at larger energies.

The most plausible resolution is that black holes states appear in
both parity channels, leading to a cancellation if the spectral
functions are the same. As discussed, this can be the case whether
or not $[H, \PP] = 0$.

This conclusion is quite general, and does not depend in detail on
the matter content of the model. It is straightforward to
generalize this result to other discrete symmetries, or even to
continuous symmetries. If the symmetry can be applied to the
low-energy amplitude (i.e. at energies small compared to $M_{\rm
Planck}$, where quantum gravity is negligible), then the
corresponding dispersion relation requires that it apply to the
black hole production process in that channel. Hence we conclude
that black holes exist in each of the channels related by the
symmetry.

%\vskip 1 in
\newpage

\centerline{\bf Acknowledgments}

The author thanks N. Deshpande, R. Hwa, L. Krauss, J. Polchinski,
D. Soper and M. Wise for discussions. This work was supported in
part under DOE contract DE-FG06-85ER40224 and NSF contract
PHY99-07949. The author acknowledges the hospitality of the Kavli
Institute for Theoretical Physics at UC Santa Barbara, where this
work was completed.

%\newpage

\end{document}